\documentclass[epj]{svjour}

\usepackage{amsmath}
\usepackage{epsfig}

\newcommand{\beq}{\begin{equation}}
\newcommand{\eeq}{\end{equation}}
\newcommand{\bea}{\begin{eqnarray}}
\newcommand{\eea}{\end{eqnarray}}
\newcommand{\Bx}{x_{bj}}
\begin{document}

\title{DVCS on nuclei: Observability and Consequences}
\author{A.~Freund$^a$ and M.~Strikman$^b$}
\institute{$^a$Institut f{\"u}r Theoretische Physik, 
Universit{\"a}t Regensburg, D-93040 Regensburg, Germany\\
$^b$Department of Physics, The Pennsylvania State University, 
University Park, PA  16802, USA}

\date{\today}
%\medskip
%\noindent Keywords: Generalized parton distributions, deeply virtual 
%Compton scattering

\abstract{In this paper, we discuss the feasibility of measuring
deeply virtual Compton scattering (DVCS) on nuclei in a collider
setting, as for example, the planned high-luminosity
Electron-Ion-Collider (EIC). We demonstrate that employing our recent
model for nuclear generalized parton distributions (nGPDs), the
one-photon unpolarized DVCS cross section as well as the azimuthal- and spin
asymmetry are of the same size as in the proton case. This
will allow for an experimental extraction of nuclear GPDs with high
precision shedding new light not only on nuclear shadowing at small $x_{bj}$
but also on the interplay of shadowing and nuclear enhancement at $x_{bj}\sim0.1$.
\PACS{{11.10.Hi} {11.30.Ly} {12.38.Bx}}}

\maketitle

\section{Introduction}

Recent experimental studies of deeply virtual Compton scattering
(DVCS) on protons at HERA \cite{ZEUS,H1} have reached a $x_{bj},Q^2$
range where they now provide a sensitive test of current models of
generalized parton distributions (GPDs) in nucleons \cite{FM,FMS}
which encode the three dimensional structure of hadrons
\cite{afgpd3d,bemunew,diehl3d,ji3dnew}.  In particular, it was shown
that the NLO predictions are very sensitive to the gluon GPD inside
the proton \cite{FM,FMS}.

Recently, we performed \cite{AFMS} studies of nuclear GPDs starting
with the nuclear parton densities modeled using the leading twist
theory of nuclear shadowing (see \cite{guide} and references therein).
We used the same approximations for connecting GPDs and forward parton
densities as were used in the proton case in \cite{FMS} which led
to a good description of the current world DVCS data.
 
When applied to the calculation of nuclear DVCS amplitudes, we
observed that the nuclear modifications were up to $50\%$ larger for
the imaginary part of the DVCS amplitude than for the equivalent
inclusive case. This is due to the sensitivity of the imaginary part
of the DVCS amplitude to larger longitudinal distances compared to the
inclusive amplitude for the same $x_{bj}$ \cite{afgpd3d}. This
enhanced sensitivity makes the imaginary part an ideal tool to study
high gluon density effects like saturation in nuclear DVCS at values
of $x_{bj}$ where these effects normally would not occur in inclusive
reactions (see \cite{afgpd3d} for a detailed discussion). At the same
time, we found a surprisingly large modification of the real part of
the DVCS amplitude extending to the region of $x_{bj}\sim0.1$, where
nuclear effects for the quark distributions are very small.
Technically, this occurred due to a delicate interplay between
shadowing corrections for quarks and nuclear enhancements for gluons
in this region (see \cite{AFMS} for details). This, however, implies
that the real part of the nuclear DVCS scattering amplitude is
extremely sensitive to different physical phenomena not present in a
nucleon occurring at longitudinal distances which are orders of
magnitude different (see for example \cite{afgpd3d} for a detailed
discussion). In summary, nuclear DVCS observables will allow us to
study the modification or deformation of the three dimensional
distribution of particle correlations encode in nGPDs due to a nuclear
environment. This will enable us to shed new light on the dynamical
interplay of highly complex bound hadronic systems.

In this paper, we will address the question of the feasibility of
experimentally testing our predictions of nuclear modifications for
GPDs. The main difficulty inherent to coherent processes with nuclei
is their being strongly dominated by scatterings at $-t\leq 3/R_A^2$ with
$R_A^2$ the mean square nuclear radius. Detection of the scattered
nuclei in this case is impossible in collider kinematics and would be
a very challenging task for a dedicated fixed target experiment.  The
current HERMES DVCS experiment \cite{hermesnuc} sums over the coherent
and incoherent contributions with the incoherent term giving a
significant contribution \cite{Guzey} (see also \cite{muki} for an
extensive study of how coherent DVCS depends on the spin of the target
as well as \cite{pirenuc}).  At the same time, it is pretty
straightforward for an experiment at the Electron-Ion Collider (EIC)
planned at BNL \cite{Whitepaper} to select the process $eA\to e\gamma A$ to
which both the QED Bethe-Heitler and DVCS process contribute.

In consequence, this paper will address the question what
$t$-integrated observables can be used to study nuclear DVCS and,
therefore, the physics of nuclear modifications to the three
dimensional distribution of particle correlations in nuclei.

It is worth emphasizing that the $t$-dependence of the DVCS amplitude
for $200\geq A \geq16 $ and small $x\leq 0.03\cdot(200/A)^{1/3}$ is predominatly
given by the nuclear body form factor and hence it is not of
particular interest. At the same time, knowing this $t$-dependence and
using information about the $t$-integrated amplitudes for a range of
nuclei will make it possible to reconstruct the DVCS amplitude as a
function of the nuclear thickness, $T(b)=\int_{-\infty}^{+\infty} \rho_A(z,b)dz$,
hence reaching a nuclear optical thickness which is about 1.5 times
larger than the average heavy nucleus thickness.

The paper is organized as follows. In Section \ref{summary} we
summarize the necessary formulae for the relevant observables: the
one-photon cross section, the azimuthal angle asymmetry (and
equivalent charge asymmetry) due to the interference of DVCS and
Bethe-Heitler amplitudes, as well as the single spin asymmetry due to
the longitudinal polarization of the electron beam (the EIC design
includes running with highly polarized electron beams). We will also
discuss the issue of the $t$-dependence.  In section \ref{results} we
give our numerical predictions for four spin zero nuclei
(O-16,Ca-40,Pd-110 and Pb-208) including kinematic twist-3 effects in
leading order (LO) (a detailed QCD analysis of kinematic twist-3
effects on the nucleon for the EIC including QCD evolution was
presented in \cite{ftw3}). Since the LO twist-3 DVCS amplitudes,
modulo the dynamic twist-3 contributions, are totally expressible
through twist-2 GPDs \cite{tw3proton}, the twist-3 nuclear DVCS
amplitudes can be directly computed from the nuclear twist-2 GPDs we
employ here. We demonstrate that it would be feasible to observe the
single spin asymmetry at the EIC. The observation of the angular
asymmetry, though it is large at the EIC, will require a very good
energy and momentum resolution of the electron and photon detectors.
An option for a positron source at the EIC would enable the easier
measurement of the charge asymmetry containing complementary
information to the azimuthal angle asymmetry. The observation of the
DVCS contribution to the total cross section would require a
measurement at the $10\%$ accuracy level. It is also demonstrated that
the twist-3 effects in nuclei are much smaller than in the nucleon and
completely negligible. We conclude with a summary where we argue that
DVCS measurements with nuclei will be an important component of the
EIC program in particular for the studies of the color transparency
phenomenon in vector meson production.

\section{Summary of DVCS formulas}
\label{summary}

\begin{figure} 
\centering 
\mbox{\epsfig{file=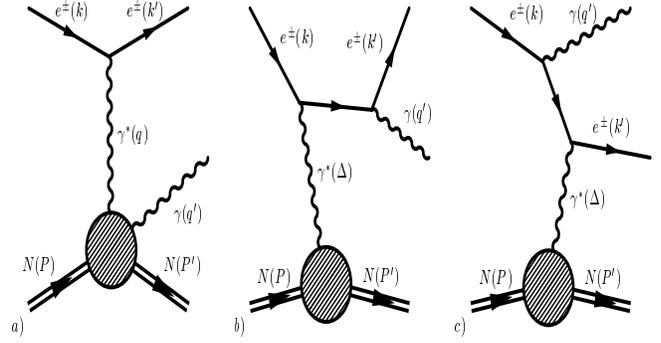,width=8.5cm,height=4.5cm}} 
\caption{a) DVCS graph, b) BH with photon from final state lepton and
  c) with photon from initial state lepton.}
\label{dvcspic} 
\end{figure}

The lepton level process, $l(k) ~A(P) \to l(k')
~A(P')~\gamma(q')$, receives contributions from each of the graphs
shown in Fig.\ \ref{dvcspic}. This means that the cross section will
contain a pure DVCS-, a pure BH- and an interference term.

\begin{figure}
\centering
\mbox{\epsfig{file=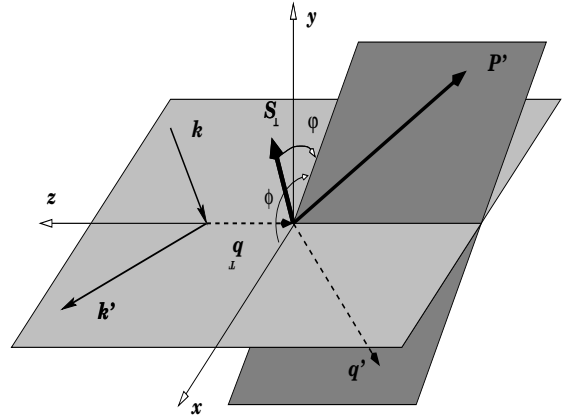,width=8.5cm,height=5.5cm}}
\caption{
  The kinematics of the leptoproduction in the target rest frame.}
\label{frame}
\end{figure}

We choose to work in the target rest frame given in, for example,
\cite{abmuki} (see Fig.~\ref{frame}), where the positive $z$-direction
is chosen along the three-momentum of the incoming virtual photon. The
incoming and outgoing lepton three-momenta form the lepton scattering
plane, while the final state nucleus and outgoing real photon define
the nucleus-photon scattering plane. In this reference frame the
azimuthal angle of the scattered lepton is $\phi_l = 0$, while the
azimuthal angle between the lepton plane and the final state nucleus
momentum is $\phi_A = \phi$. When the hadron is transversely polarized
(within this frame of reference) $S_\perp = (0, \cos {\mit\Phi}, \sin
{\mit\Phi}, 0)$ and the angle between the polarization vector and the
scattered hadron is given by $\varphi = {\mit\Phi} - \phi_A$. The four vectors are
$k = (E, E \sin\theta_l, 0, E \cos\theta_l )$, $q = (q^0, 0, 0,-|q^3|)$. Other
vectors are $P = (M_A, 0, 0, 0)$ and $P' = (E_A',
|\mbox{\boldmath$P'$}| \cos\phi \sin\theta_A, | \mbox{\boldmath$P'$}| \sin\phi
\sin\theta_A, |\mbox{\boldmath$P'$}| \cos\theta_A)$.  The longitudinal part of
the polarization vector is $S_{\rm LP} = (0, 0, 0, \Lambda)$. The relevant
Lorentz-invariant variables for DVCS are then:
\begin{eqnarray} 
\xi_A = \frac{Q^2}{2 {\bar P} \cdot {\bar q}} \, , 
{\bar {\cal Q}}^2 = -{\bar q}^2, 
t = \Delta^2 = (P-P')^2 \, ,
y = \frac{P\cdot q}{P\cdot k}\, , 
\nonumber 
\end{eqnarray} 
where ${\bar P} = (P+P')/2$, ${\bar q} = (q+q')/2$ and which are
related to the experimentally accessible variables, $\zeta \equiv x_{bj} =
-Aq^2/2P\cdot q = A\zeta_A$ and $Q^2 = - q^2$, used throughout this paper, via
\begin{align} 
&{\bar {\cal Q}}^2 = \frac{1}{2}Q^2\left(1+\frac{t}{Q^2}\right)\approx \frac{1}{2}Q^2\nonumber\\
&\xi_A = \frac{\zeta_A\left(1+\frac{t}{2 Q^2}\right)}{2-\zeta_A\left(1-\frac{t}{Q^2}\right)}\approx 
\frac{\zeta}{2A-\zeta} \, . \label{qbar} 
\end{align}
Note that $t$ has a minimal value given by
\begin{eqnarray}
\label{Def-tmin}
-t_{\rm min}^2
=  Q^2
\frac{2A(A - \Bx) \left(1 - \sqrt{1 + \epsilon^2}\right) + A^2\epsilon^2}
{4\Bx (A - \Bx) + A^2\epsilon^2}
\, .
\end{eqnarray}
where $\epsilon^2= 4M_N^2\Bx^2/Q^2$.

The corresponding five-fold differential cross section for a spin-$0$
target per nucleon is given by \cite{abmuki} to twist-3 accuracy
neglecting transversity contributions:
\begin{eqnarray}
\label{WQ}
\frac{1}{A}\frac{d\sigma}{d\Bx dy d|t| d\phi d\varphi}
=
\frac{\alpha^3  \Bx y } {16 \, \pi^2 \,  Q^2 \sqrt{1 + \epsilon^2}A^3}
\left| \frac{\cal T}{e^3} \right|^2 \, .
\end{eqnarray}
Note that we give our nuclear amplitudes per nucleon \cite{AFMS}. In
order to obtain the correct scaling of the cross section with $A$, the
DVCS amplitude has to be multiplied by $A$. This leads to an extra
factor of $A^2$ in the cross section which cancels two of the powers
of $A$ in Eq.~(\ref{WQ}) to finally yield the correct overall
power of $1/A$.

The square of the amplitude receives contributions from pure DVCS
(Fig.\ 1a), from pure BH (Figs. 1b, 1c) and from their interference
(with a sign governed by the lepton charge), \bea |{\cal T}|^2 =
|{\cal T}_{DVCS}|^2 + {\cal I} + |{\cal T}_{BH}|^2
\label{tdef} 
\eea 
where the individual terms are given by
\begin{eqnarray}
\label{Par-BH}
&&|{\cal T}_{\rm BH}|^2
= - \frac{e^6F^2(t)}
{\Bx^2 y^2 (1 + \epsilon^2)^2 t\, {\cal P}_1 (\phi) {\cal P}_2 (\phi)}\times
\nonumber\\
&&\left[
c^{\rm BH}_0
+  \sum_{n = 1}^2
c^{\rm BH}_n \, \cos(n\phi)\right]
\\
\label{AmplitudesSquared}
&& |{\cal T}_{\rm DVCS}|^2
=
\frac{e^6}{\Bx y^2 {\cal Q}^2}\times\nonumber\\
&&\Big[
c^{\rm DVCS}_0
+
\left[c^{\rm DVCS}_1 \cos (\phi) + s^{\rm DVCS}_1 \sin (\phi)\right]
\Big]
\\
\label{InterferenceTerm}
&&{\cal I}
= \frac{\pm e^6F(t)}{\Bx^2 y^3 t {\cal P}_1 (\phi) {\cal P}_2 (\phi)}\nonumber\\
&&\Big[
\frac{t}{Q^2}c_0^{\cal I}
+ \sum_{n = 1}^2
\left[c_n^{\cal I} \cos(n \phi) +  s_n^{\cal I} \sin(n \phi)\right]
\Big]
\end{eqnarray}
where the $+/-$ sign in the interference stands for a
negatively/positively charged lepton and $F(t)$ is the nuclear form
factor.

The $c_n$'s and $s_n$'s are the Fourier coefficients of the $\cos(n\phi)$
and $\sin(n\phi)$ terms. These coefficients are given as combinations of
the real and imaginary part of the unpolarized DVCS amplitudes ${\cal
  H}$ (for the $c^{\cal I}$'s or $s^{\cal I}$'s) or the squares of the
afore mentioned DVCS amplitude (for the $c^{\rm DVCS}$'s or $s^{\rm
  DVCS}$'s). The exact form is given in \cite{abmuki} and does not
have to be repeated here. The computation of the DVCS amplitudes and
the necessary model assumptions were extensively discussed in
\cite{AFMS} and will not be repeated here.  The precise form of the BH
propagators ${\cal P}_{1,2}(\phi)$ which induces an additional
$\phi$-dependence, besides the $\cos(n\phi)$ and $\sin(n\phi)$ terms, and which
can mock $\cos(n\phi)$ and $\sin(n\phi)$ dependences in certain observables,
can also be found in \cite{abmuki}.  Note that in order to avoid
collinear singularities occurring through the coincidence of the
outgoing photon with the incoming lepton line in ${\cal P}_{1,2}(\phi)$
we need to constrain $y$ according to 
\bea y \leq y_{\rm col} \equiv \frac{Q^2
  + t}{Q^2 + \Bx t}.  
\eea 
in order to avoid an artificially enhanced
BH contribution. This limit is only of practical relevance for fixed
target experiments at very low energies. Collider experiments do not
have any meaningful statistics for exclusive processes at very large
$y\simeq O(1)$.

The DVCS observables we will discuss later on are based on a less
differential cross section than the five-fold one in Eq.~(\ref{WQ}).
The reason for this is first that the cross section in Eq.~(\ref{WQ})
is frame dependent since the azimuthal angles $\phi$ and $\varphi$ are not
Lorentz invariants and hence, they will be integrated out. Secondly,
since a $t$-distribution in nuclei will be impossible to measure, we
also integrate out $t$, however, with experimentally sensible cuts as
will be discussed later. In consequence, our observables will be based
on only a two-fold differential cross section.  In the following, we
will concentrate on the Single Spin Asymmetry (SSA), the Charge
Asymmetry (CA) and the azimuthal angle asymmetry (AAA) defined in accordance
with experiments the following way:
\begin{align}
&SSA = \frac{2 \int_0^{2\pi}d\phi ~\sin(\phi)(d\sigma^{\uparrow}-d\sigma^{\downarrow})}{\int_0^{2\pi}d\phi~(d\sigma^{\uparrow}+d\sigma^{\downarrow})} \, ,
\label{defssa}\\
&CA =  \frac{2\int_0^{2\pi}d\phi ~\cos(\phi)(d\sigma^{+}-d\sigma^{-})}{\int_0^{2\pi}d\phi~(d\sigma^{+}+d\sigma^{-})} \, , 
\label{defca}\\
&AAA
=\frac{\int^{\pi/2}_{-\pi/2} d\phi (d\sigma-d\sigma^{BH}) - \int^{3\pi/2}_{\pi/2}d\phi (d\sigma-d\sigma^{BH})}{\int^{2\pi}_{0} d\phi d\sigma}\, .
\label{aaadef}
\end{align}
Here $d\sigma^{\uparrow}$ and $d\sigma^{\downarrow}$ refer to the two fold differential cross
sections $d\sigma/d\Bx dQ^2$ with the lepton polarized along or against its
direction of motion, respectively; $d\sigma^{+}$ and $d\sigma^{-}$ are the
unpolarized differential cross sections for positrons and electrons,
respectively and $d\sigma^{BH}$ refers only to the pure BH cross section.

The EIC will most likely be running with electrons only and therefore
the CA will not be measured. This leaves only the SSA and AAA. The
experimental problem or challenge with the AAA, however, is that it
requires either a very good detector resolution i.e. many bins in $\phi$
or an event by event reconstruction of the scattering planes. The last
statement needs a word of explanation: Eq.~(\ref{aaadef}) is
equivalent to taking the difference between the number of DVCS minus
BH events where the real $\gamma$ is above the electron scattering plane
and where it is below that plane, divided by the total number of
events. This procedure ensures that the numerator is not contaminated
by BH, which would spoil an unambiguous interpretation of the
observable in terms of the real part of DVCS amplitudes. Also, the
only difference between Eq.~(\ref{defca}) and (\ref{aaadef}) is due to
the additional interference term in the denominator of
Eq.~(\ref{aaadef}) and a twist-2$\times$twist-3 contribution in the DVCS
squared part which are both small in a collider setting compared with
the leading contribution. Therefore, it does not matter from a physics
point of view, whether one discusses the AAA or the CA!
  
We will also discuss the one-photon cross section $\sigma(\gamma^*A)$ at small
$\Bx$ defined through
\begin{align} 
&\frac{d^{2}\sigma (eA \to eA\gamma)}{dy dQ^2} = \Gamma~\sigma_{DVCS}(\gamma^*A\to 
\gamma A)\nonumber\\
&~\qquad~\mbox{where}~~\qquad~\Gamma = \frac{\alpha_{e.m.} (1+(1-y)^2)}{2\pi 
  y Q^2}. 
\end{align}
with
\begin{align} 
\frac{1}{A}\sigma_{DVCS}(\gamma^*A \to \gamma A) = \frac{\alpha^2\Bx^2\pi}{Q^4{\cal B}A}|{\cal 
  T}_{DVCS}|^2|_{t=0},
\label{sigonephot} 
\end{align} 
and where ${\cal B}$ stems from the $t$-integration and will depend on
both our cut-off in $t$ and the model of the $t$-dependence we will
choose for the GPDs. Furthermore, all higher twist effects are
neglected in this quantity.

Since we have already extensively discussed our nuclear GPD model in
\cite{AFMS}, we will not repeat this discussion here, except to remind
the reader that it is based on our successful GPD parameterization for
the nucleon \cite{FMS} together with available nuclear shadowing
parameterizations. For our discussion here, we will stick to the
shadowing parameterization of \cite{guide} together with the CTEQ6
parameterization \cite{cteq6} as our necessary forward input
distribution. Furthermore, we chose to model the nuclear
$t$-dependence based on the parameterizations of the two component
Fermi models employed in \cite{guide}. We also assume that the
$t$-dependence factorizes from the GPD and thus also from the
amplitude. This is, as in the case of the nucleon, not justified at
all (see \cite{ftw3} for an extended discussion on this subject for
the nucleon case). However, we are considering the cross section
integrated over $t$, which is dominated by $-t\leq 3/R_A^2$ i.e. $-t\leq
0.01 GeV^2$ for all nuclei.  In the case of small $\Bx$ we can use our
experience of soft physics which indicates that rescattering effects
change the slope of the amplitude by at most $5\%$. For larger $\Bx$,
it also seems natural to expect non-factorizability effects to be
small for small $t$.  Hence, we expect this effect to be totally
negligible, especially when compared to other factors like nucleus
dissociation. Note also that though we perform all our calculations
with a cut on our $t$ integration at $-t_{max}=0.05~\mbox{GeV}^2$, the
contribution of $ -t \geq -t_{max}$ is negligible.

With the above discussion we have now all the tools in hand to
estimate the sizes of nuclear DVCS observables at the EIC.

\section{Results for DVCS observables}
\label{results}

For the purposes of our discussion we choose to work with the maximal
expected EIC setting for nuclei of $10~\mbox{GeV}$ electrons and
$100~\mbox{GeV}$ nuclei. The range in $\Bx$ is between $0.001-0.1$ and
the $Q^2$ range is from $3-50~\mbox{GeV}^2$ though the figures will
only show the most advantageous kinematics which does not extend to the
highest $Q^2$.

\begin{figure}  
\centering
\mbox{\epsfig{file=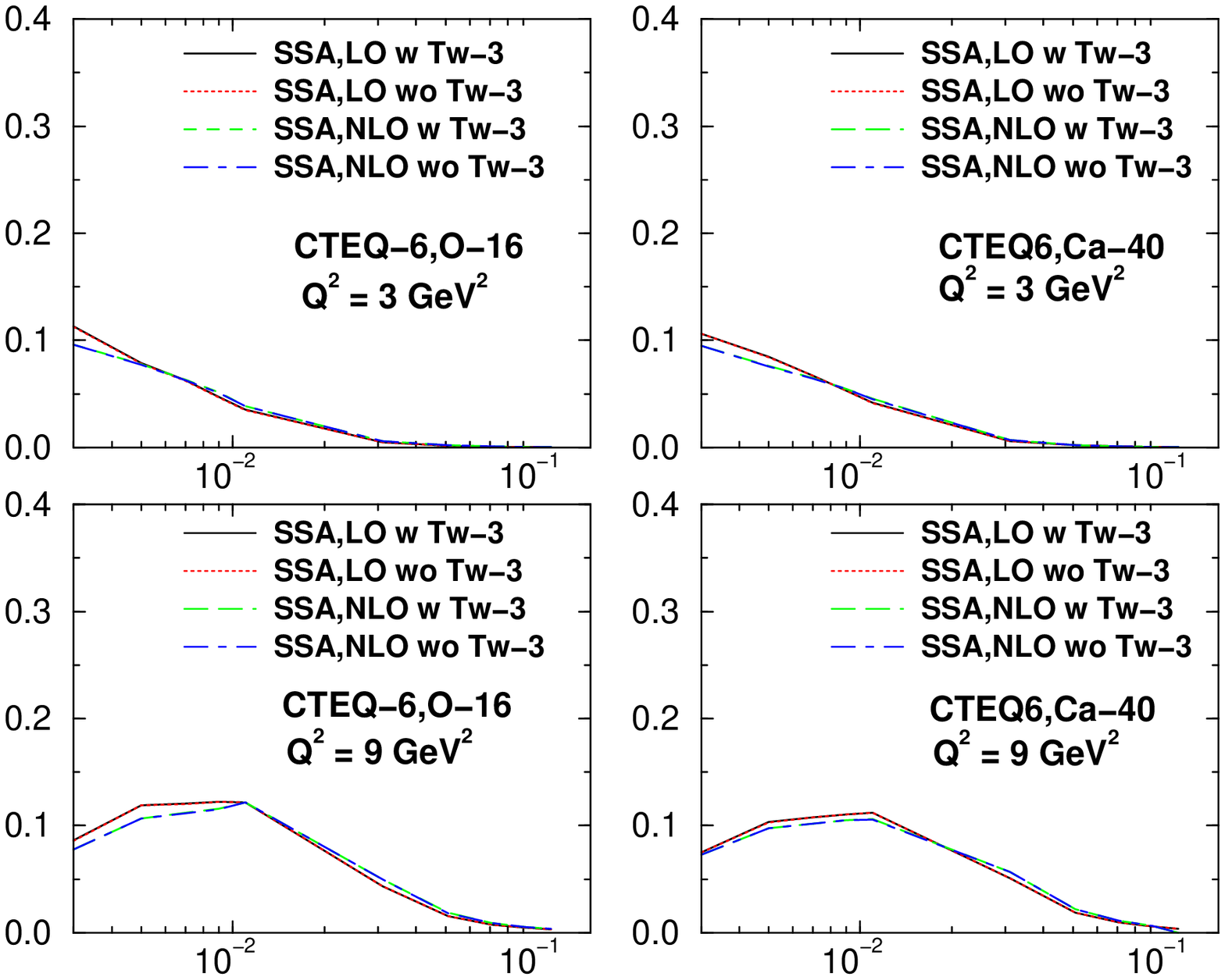,width=8.5cm,height=10.5cm}} 
\caption{$t$ integrated SSA in EIC kinematics vs. $\Bx$ for two typical values of $Q^2$ and $t_{max}= -0.05~\mbox{GeV}^2$. ``W'' stands for with and ``WO'' stands for without.}
\label{eicssaqvsx}
\vskip+0.2in
\mbox{\epsfig{file=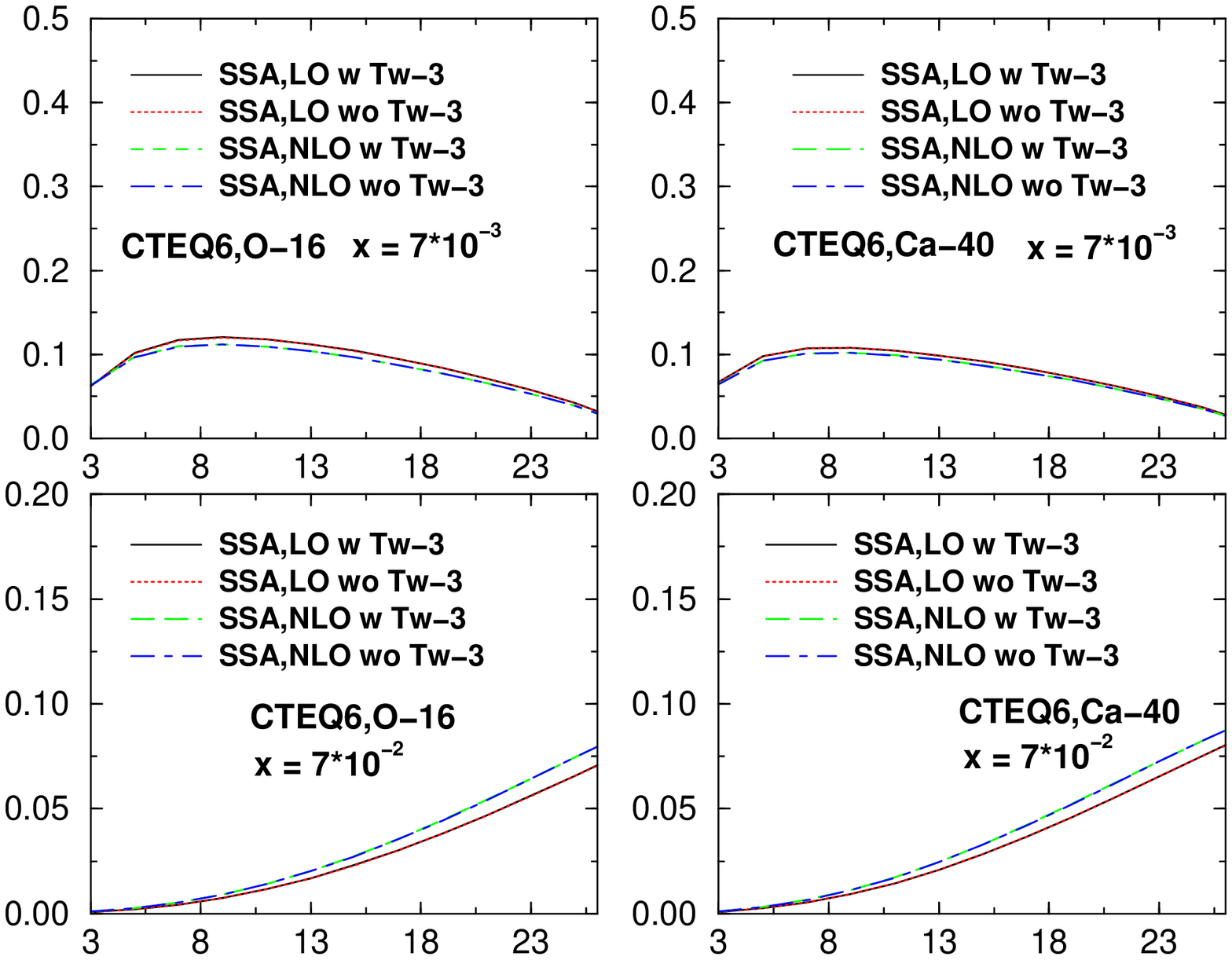,width=8.5cm,height=10.5cm}} 
\caption{$t$ integrated SSA in EIC kinematics vs. $Q^2$ for two  typical values of $\Bx$ and $t_{max}= -0.05~\mbox{GeV}^2$. ``W'' stands for with and ``WO'' stands for without.}
\label{eicssaxvsq1}
\end{figure}

\begin{figure}  
\centering
\mbox{\epsfig{file=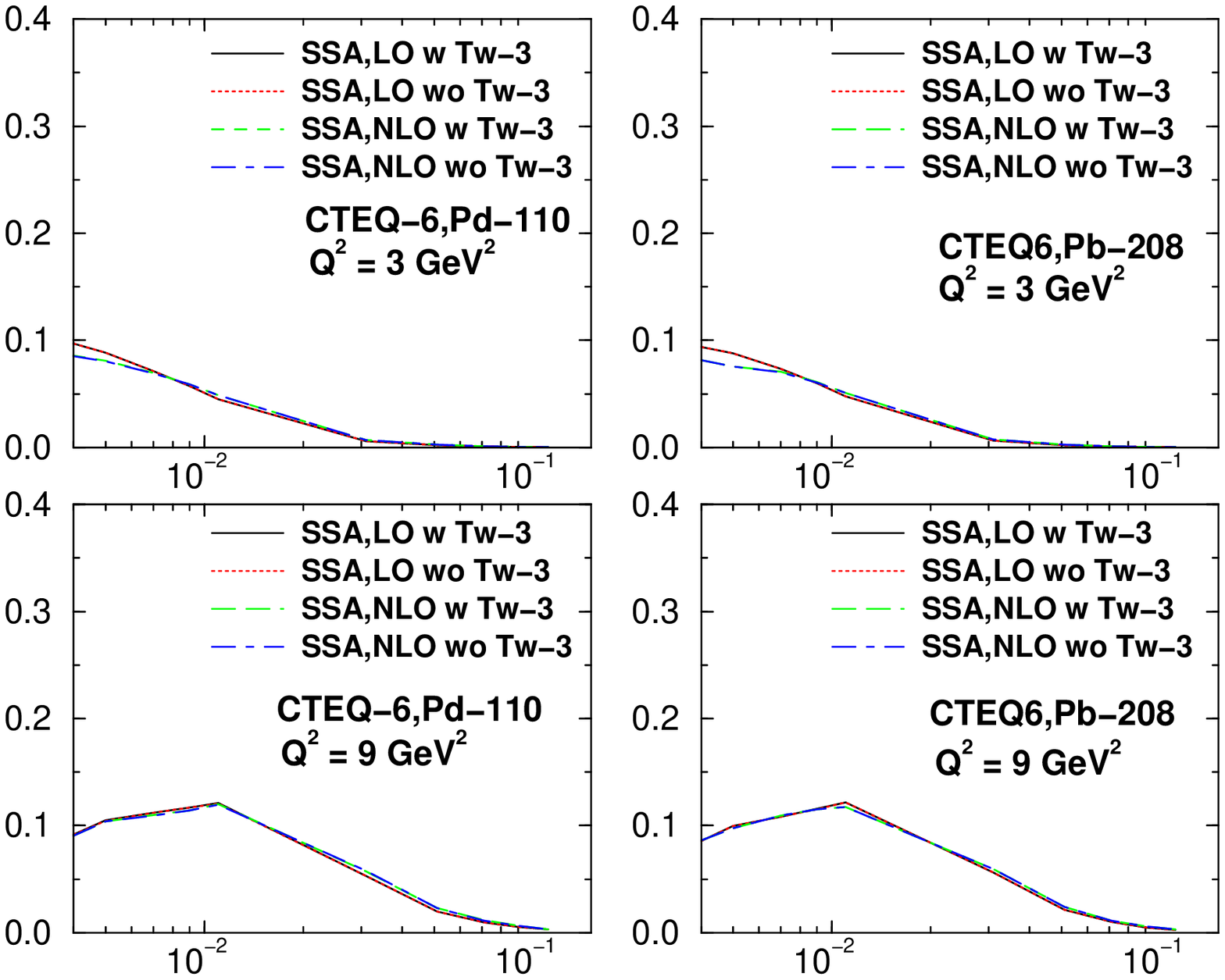,width=8.5cm,height=10.5cm}} 
\caption{$t$ integrated SSA in EIC kinematics vs. $\Bx$ for two typical values of $Q^2$ and $t_{max}= -0.05~\mbox{GeV}^2$. ``W'' stands for with and ``WO'' stands for without.}
\label{eicssaqvsx2}
\vskip+0.2in
\mbox{\epsfig{file=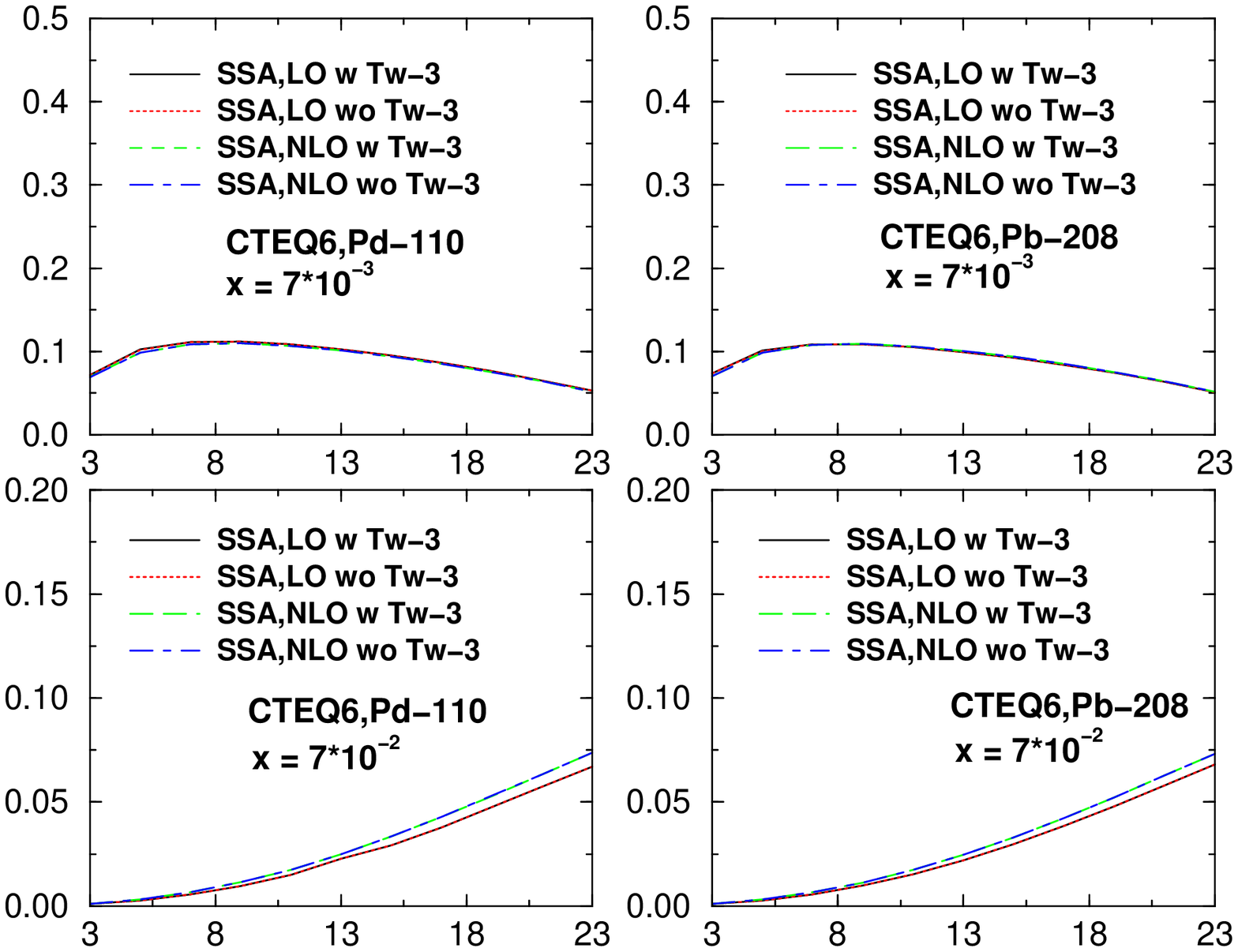,width=8.5cm,height=10.5cm}} 
\caption{$t$ integrated SSA in EIC kinematics vs. $Q^2$ for two  typical values of $\Bx$ and $t_{max}= -0.05~\mbox{GeV}^2$. ``W'' stands for with and ``WO'' stands for without.}
\label{eicssaxvsq3}
\end{figure}

Since the EIC electron beam is expected to have polarization close to
one, it would seem easiest to measure the SSA which directly probes
the imaginary part of the DVCS amplitude.  One can see from
Figs.~\ref{eicssaqvsx} -- \ref{eicssaxvsq3} that we predict the
asymmetry to be of the order of $10\%$ in a wide kinematic range which
is roughly $25-50\%$ less than in the nucleon case and nicely
demonstrates the nuclear shadowing effects in the imaginary part of
the DVCS amplitude. As a note on the side, we would like to quote our
number for the SSA for coherent DVCS integrated over $t$ for O-16 for
the average HERMES kinematics of $<Q^2> =
2.2~\mbox{GeV}^2,<\Bx>=0.09$. We find an SSA of $-0.24$ (LO) and
$-0.22$ (NLO) for O-16 compared to the HERMES SSA for Neon of
$-0.22\pm0.03\pm0.03$\cite{hermesnuc}.  Note, however, that the
preliminary HERMES data on nuclear DVCS do contain a contribution from
incoherent DVCS, since the missing mass cut of $M_X\leq 1.7 GeV$ may not
remove all incoherent contributions with associated pion production
and it definitely does not remove the contribution of the nucleus
break-up channel.

\begin{figure}  
\centering
\mbox{\epsfig{file=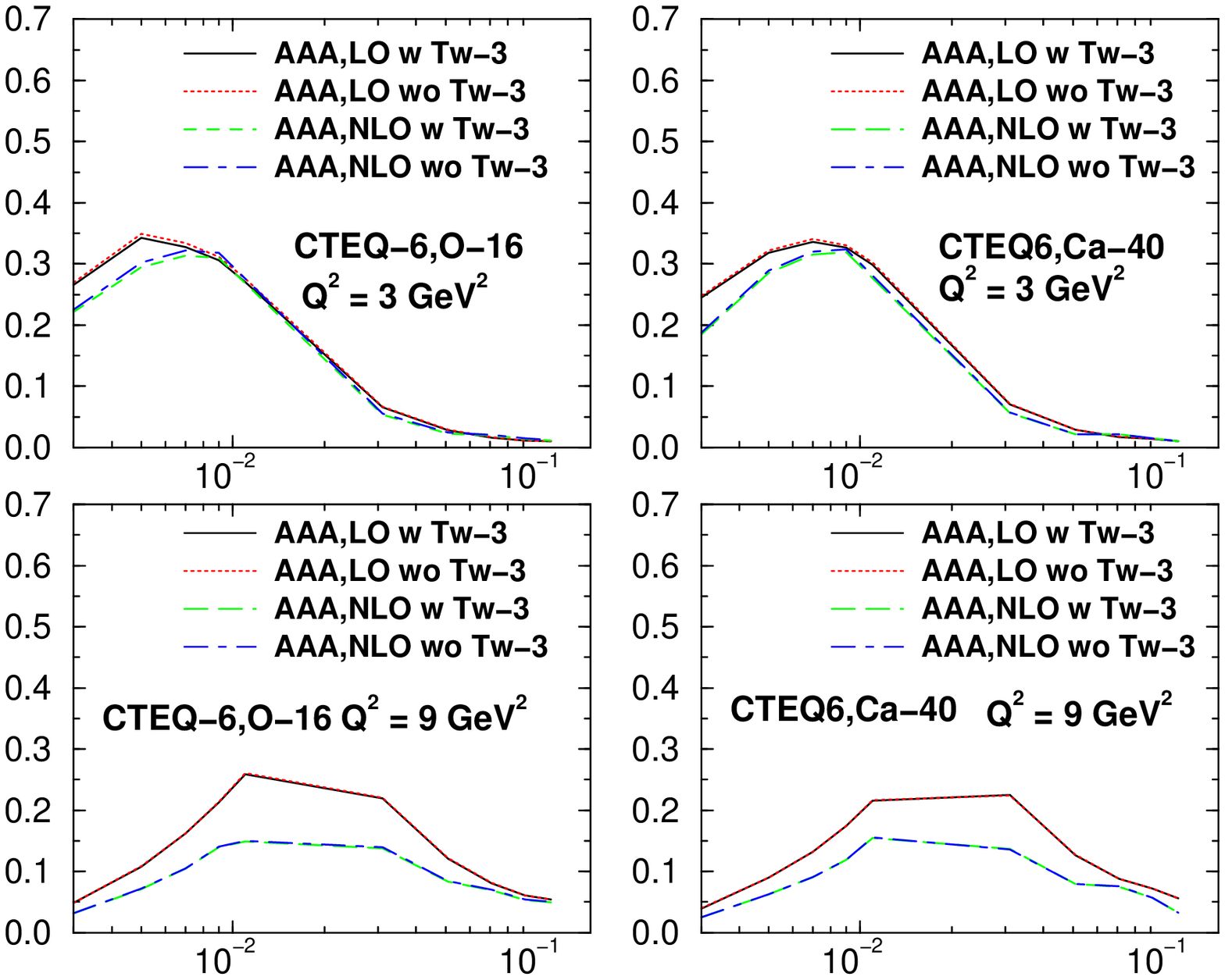,width=8.5cm,height=10.5cm}} 
\caption{$t$ integrated AAA in EIC kinematics vs. $\Bx$ for two typical values of $Q^2$ and $t_{max}= -0.05~\mbox{GeV}^2$.``W'' stands for with and ``WO'' stands for without. }
\label{eicaaaqvsx}
\vskip+0.2in
\mbox{\epsfig{file=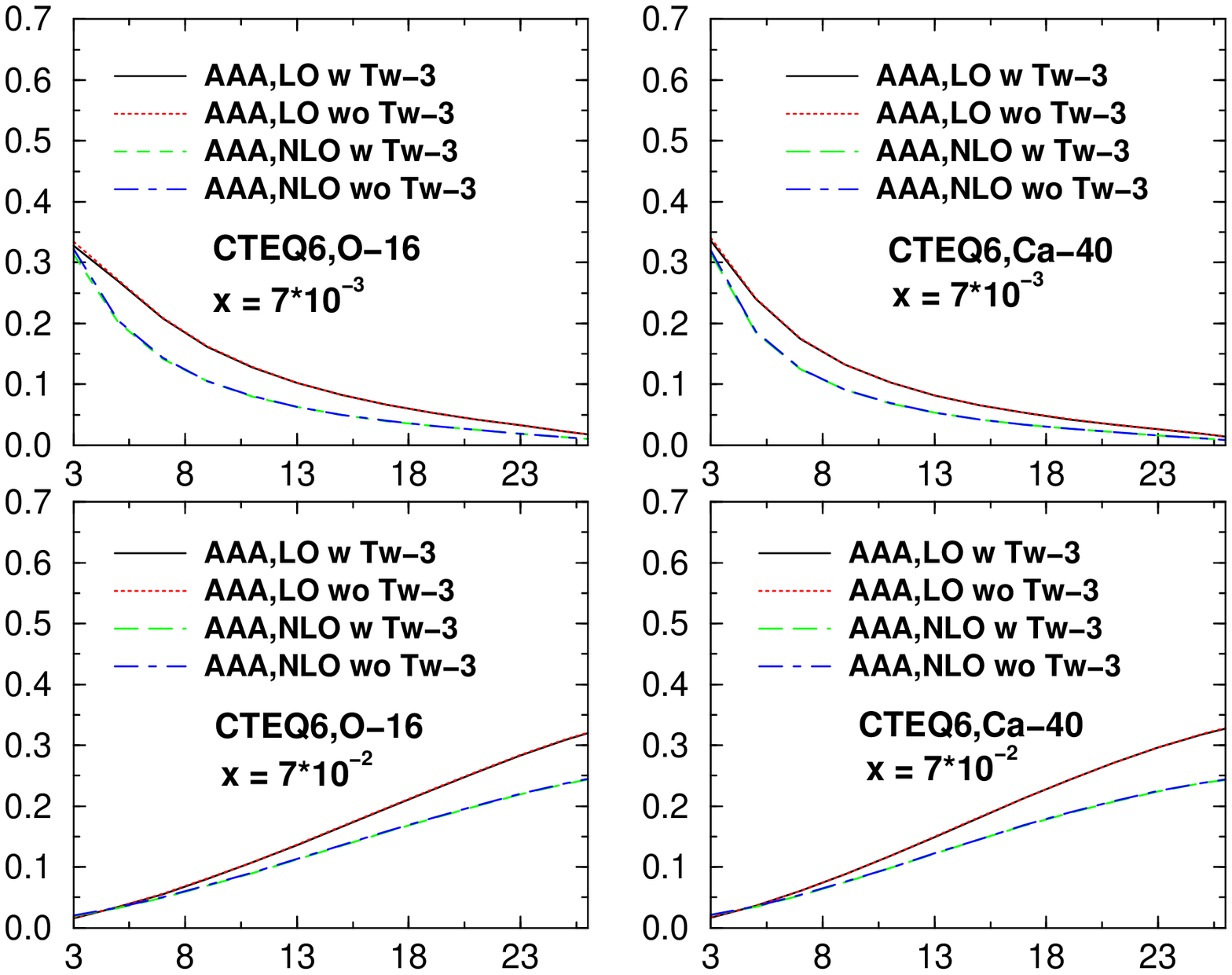,width=8.5cm,height=10.5cm}} 
\caption{$t$ integrated AAA in EIC kinematics vs. $Q^2$ for two  typical values of $\Bx$ and $t_{max}= -0.05~\mbox{GeV}^2$. ``W'' stands for with and ``WO'' stands for without.}
\label{eicaaaxvsq1}
\end{figure}

\begin{figure}  
\centering
\mbox{\epsfig{file=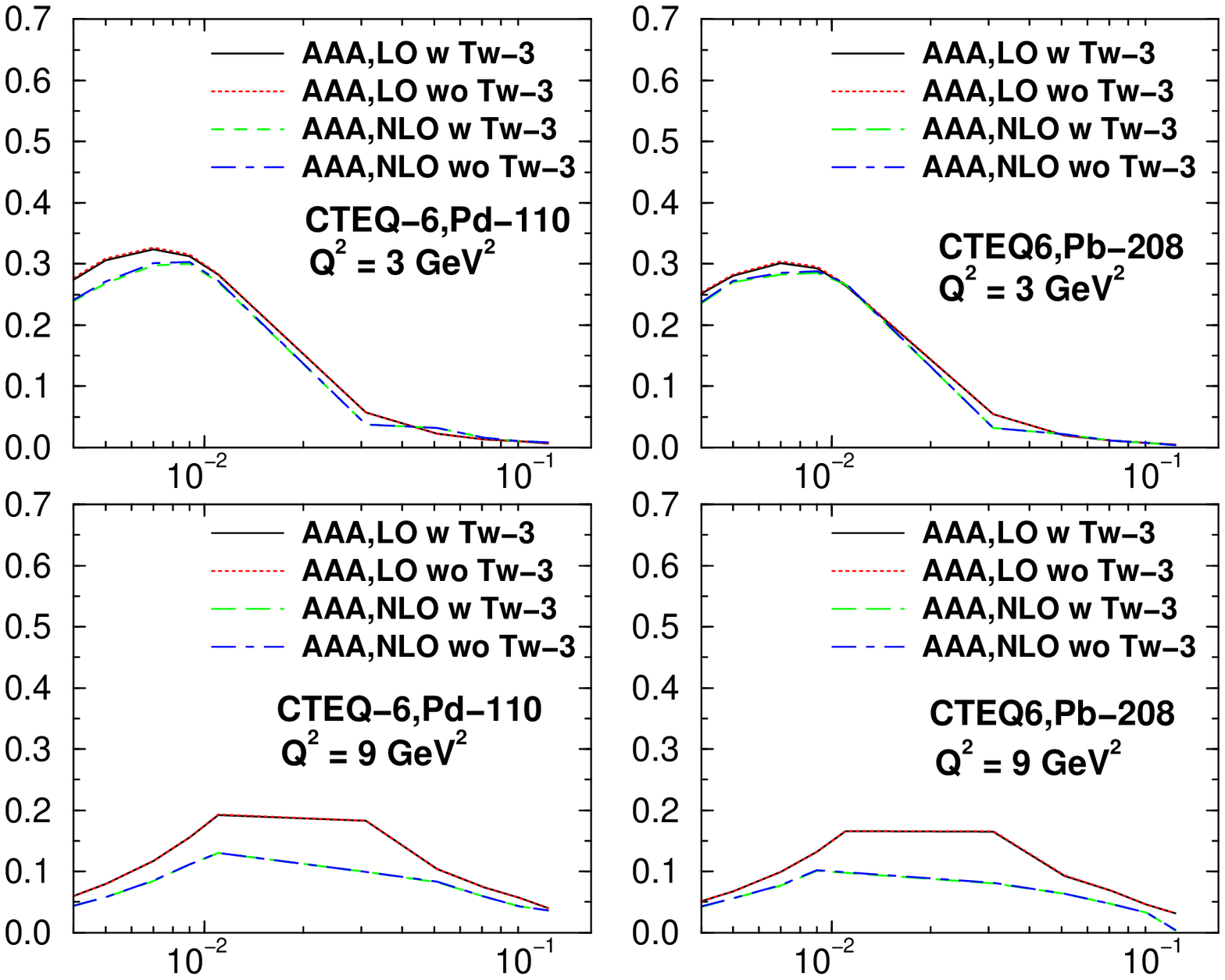,width=8.5cm,height=10.5cm}} 
\caption{$t$ integrated SSA in EIC kinematics vs. $\Bx$ for two typical values of $Q^2$ and $t_{max}= -0.05~\mbox{GeV}^2$.``W'' stands for with and ``WO'' stands for without. }
\label{eicaaaqvsx2}
\vskip+0.2in
\mbox{\epsfig{file=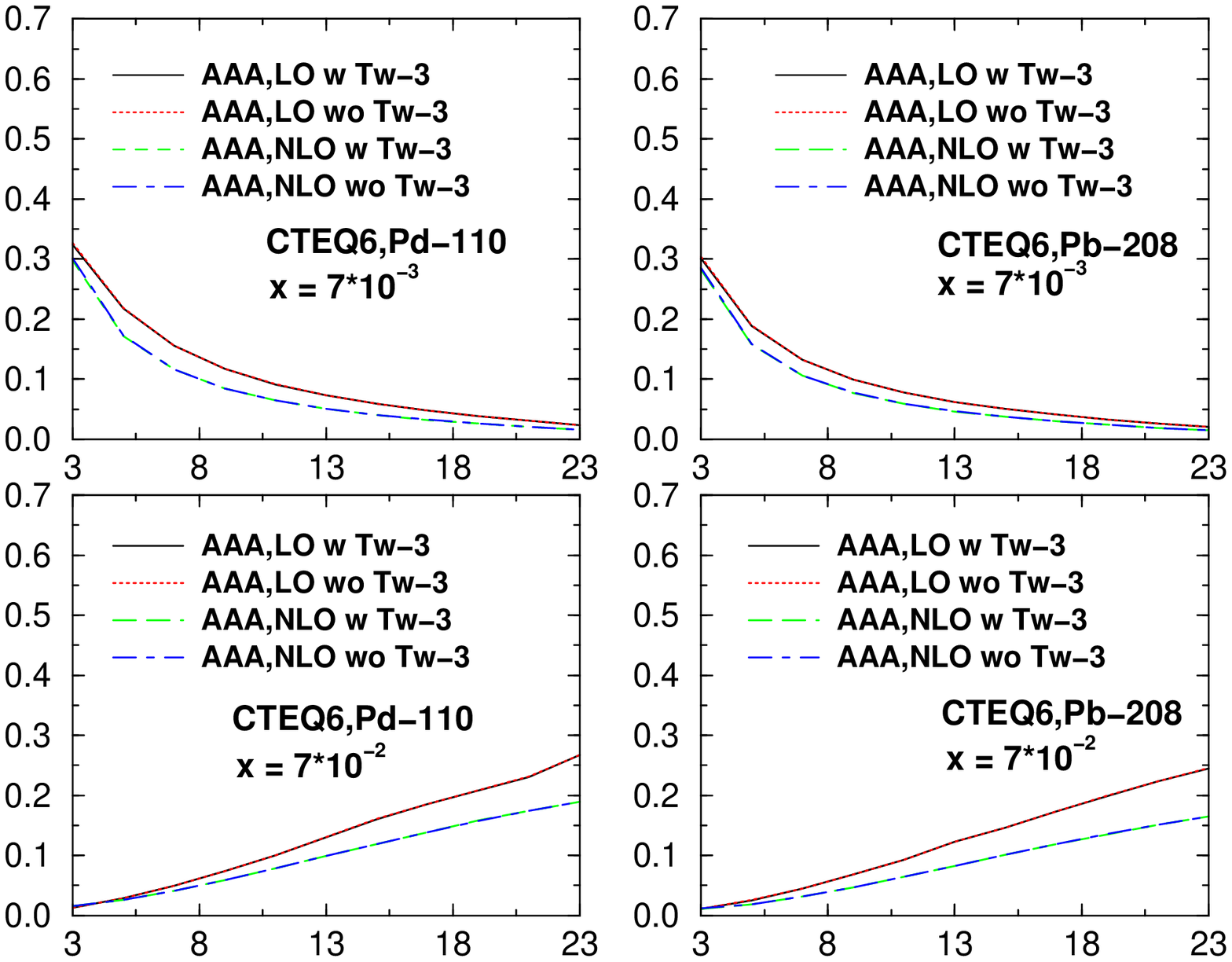,width=8.5cm,height=10.5cm}} 
\caption{$t$ integrated AAA in EIC kinematics vs. $Q^2$ for two  typical values of $\Bx$ and $t_{max}= -0.05~\mbox{GeV}^2$.``W'' stands for with and ``WO'' stands for without. }
\label{eicaaaxvsq3}
\end{figure}

We also predict a large azimuthal angular asymmetry (AAA) for all
considered nuclei, Figs.~\ref{eicaaaqvsx} -- \ref{eicaaaxvsq3}, though
a measurement of this asymmetry maybe require a very good angular
resolution. Note that the asymmetry is as large or even a bit larger
than in the nucleon case (see \cite{ftw3}). This is very surprising
given the fact that we expect a very large suppression of the real
part of the DVCS amplitude in nuclei compared to the nucleon
\cite{AFMS}. The answer to this apparent conundrum is hidden in the
details of the calculation. The numerator of the AAA receives mainly
contributions from $c^{\cal I}_0$ and $c^{\cal I}_1$ in our nuclear
GPD model for both the nucleon and nuclei. However, for the nucleon
which is spin $1/2$, there is a relative minus sign between these two
terms compared to the spin $0$ case we need to consider here, where
there is a plus sign!  This compensates for the relative suppression
as compared to the nucleon case. Furthermore, in the kinematics where
DVCS dominates over BH, the nuclear shadowing effects reduce the
denominator relative to the nucleon case which also partially
compensates for the strong suppression effect in the real part of the
nuclear amplitude. The measurement of the CA would be more advantageous
from an experimental point of view since it contains the same
information as the AAA. 

Notwithstanding the experimental difficulties, a measurement of either
the AAA or the CA would enable us to gain valuable insight into how
particle correlations i.e. the nGPDs with the three dimensional
information content are modified by the nuclear environment on very
different longitudinal distance scales.  Furthermore, we want to
emphasize here that the measurement of the real part would be
especially interesting since it is sensitive to the energy behavior of
the cross sections at smaller $\Bx$ than the ones allowed by EIC
kinematics. Hence, it is more sensitive to the proximity of the
saturation/black body limit (see discussion in \cite{FFS}).

As one can see from our calculation, the kinematic twist-3 effects are
entirely negligible for the EIC (as is similarly the case for the
nucleon \cite{ftw3}) in both the SSA and AAA. Furthermore, due to the
nuclear shadowing corrections for the gluon reducing is relative
numerical importance in the NLO amplitude, the NLO corrections for the
SSA and AAA are less than in the case of the nucleon.

\begin{figure}  
\centering
\mbox{\epsfig{file=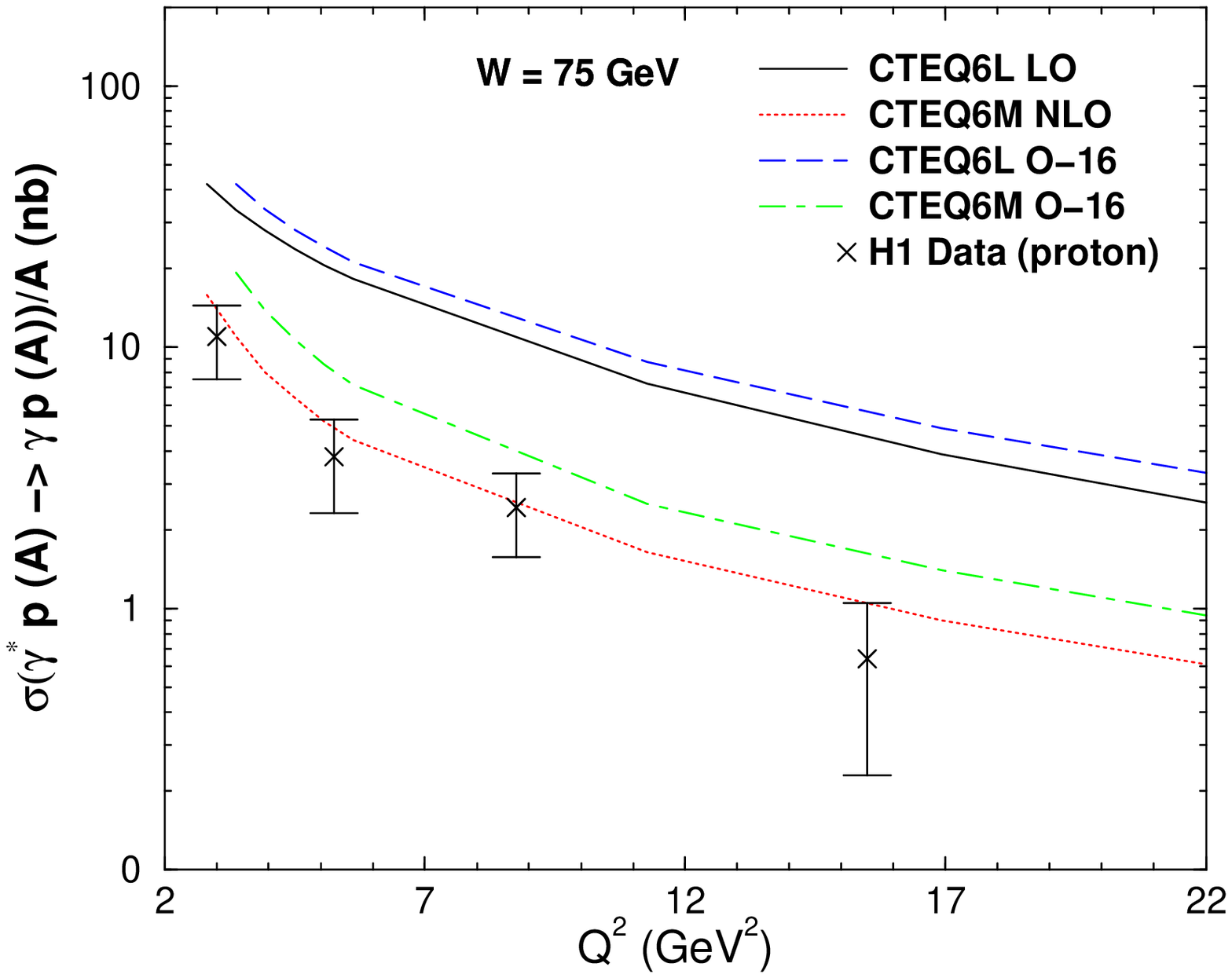,width=8.5cm,height=10.5cm}} 
\caption{One-photon cross section $\sigma(\gamma^*p)$ per nucleon for fixed $W$ vs. $Q^2$. H1 data on the nucleon is plotted in comparison within the same kinematics.}
\label{sigqo-16}
\vskip+0.2in
\mbox{\epsfig{file=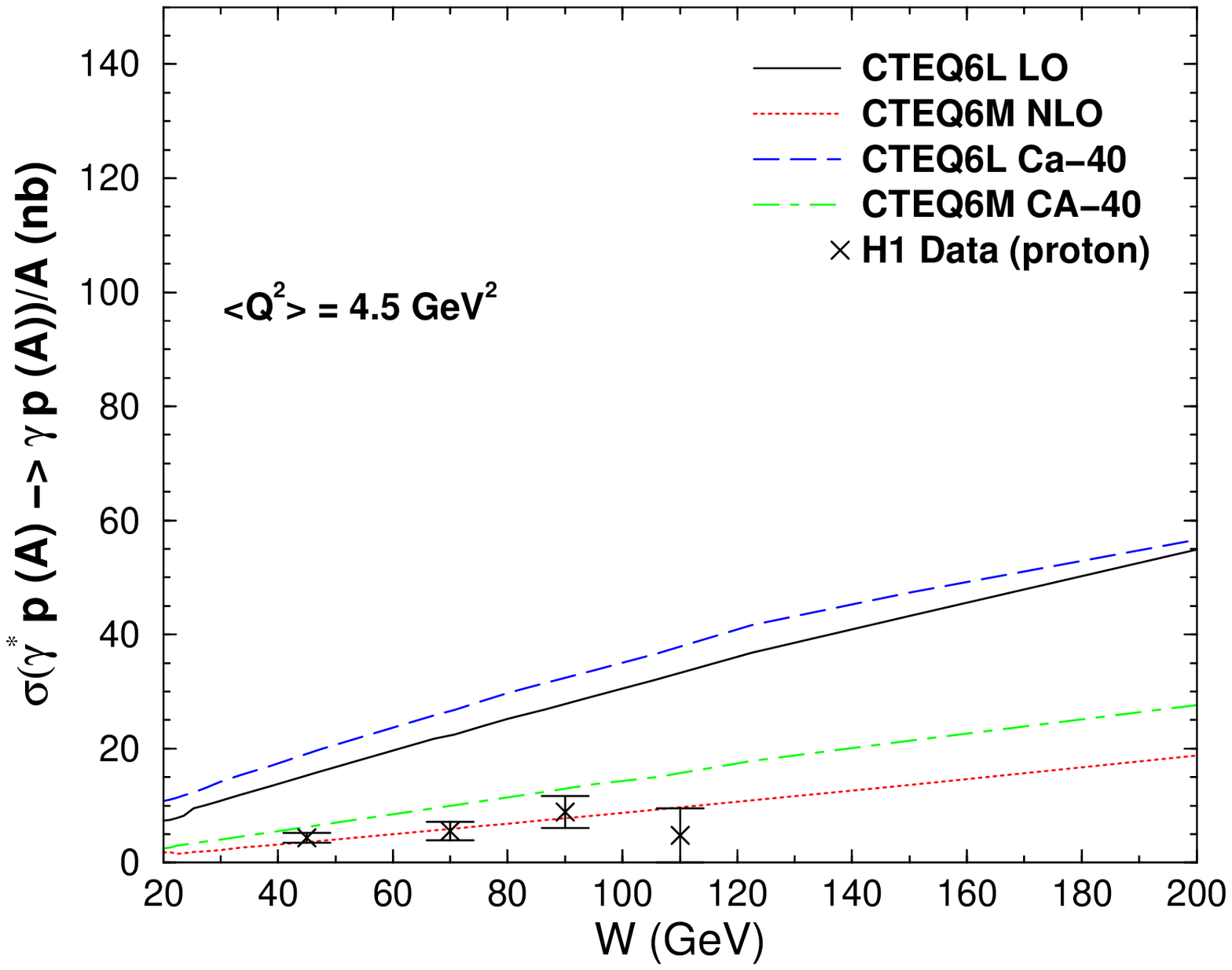,width=8.5cm,height=10.5cm}} 
\caption{One-photon cross section $\sigma(\gamma^*p)$ per nucleon for fixed $Q^2$ vs. $W$. H1 data on the nucleon is plotted in comparison within the same kinematics.}
\label{sigwo-16}
\end{figure}

\begin{figure}  
\centering
\mbox{\epsfig{file=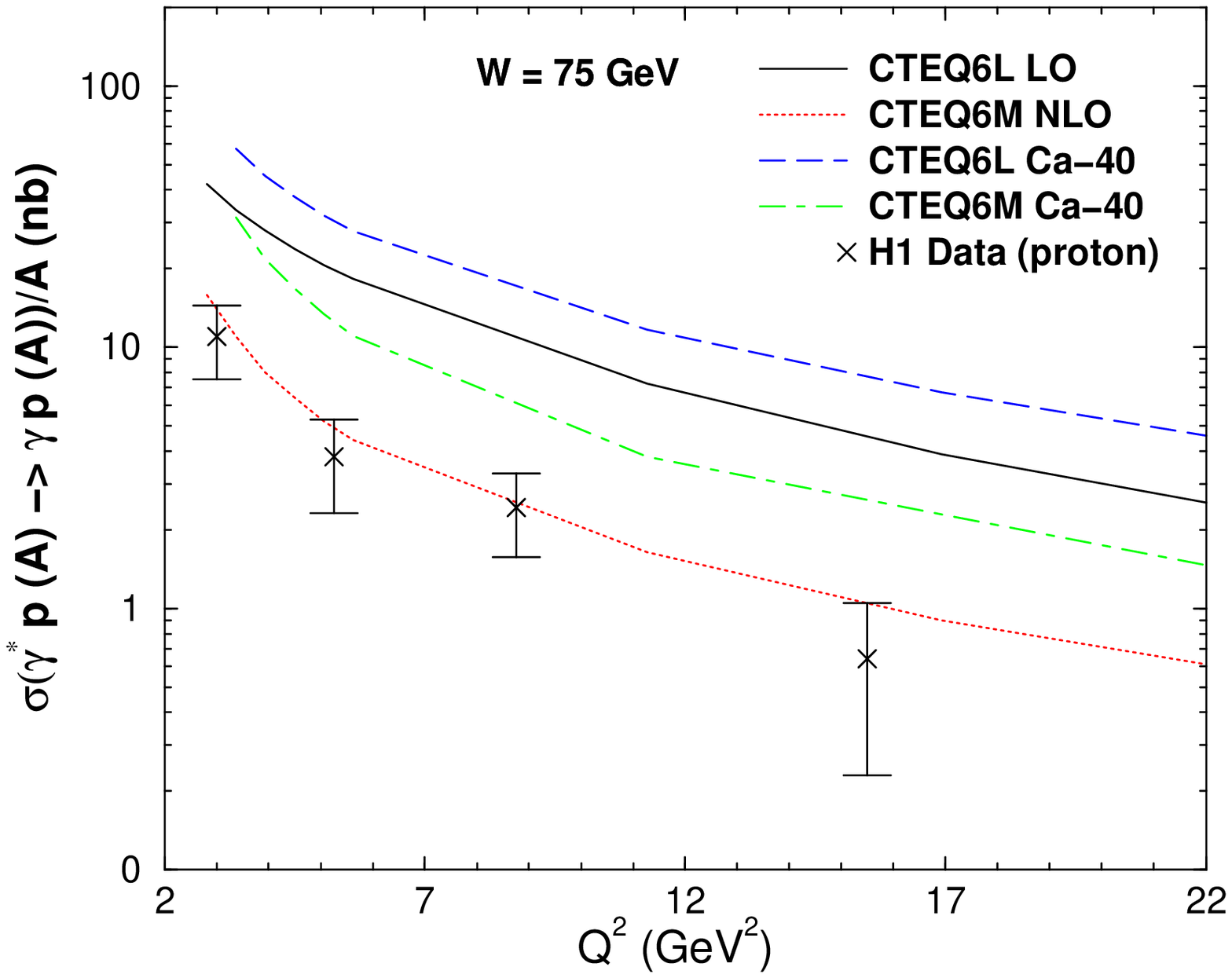,width=8.5cm,height=10.5cm}} 
\caption{One-photon cross section $\sigma(\gamma^*p)$ per nucleon for fixed $W$ vs. $Q^2$. H1 data on the nucleon is plotted in comparison within the same kinematics.}
\label{sigqca-40}
\vskip+0.2in
\mbox{\epsfig{file=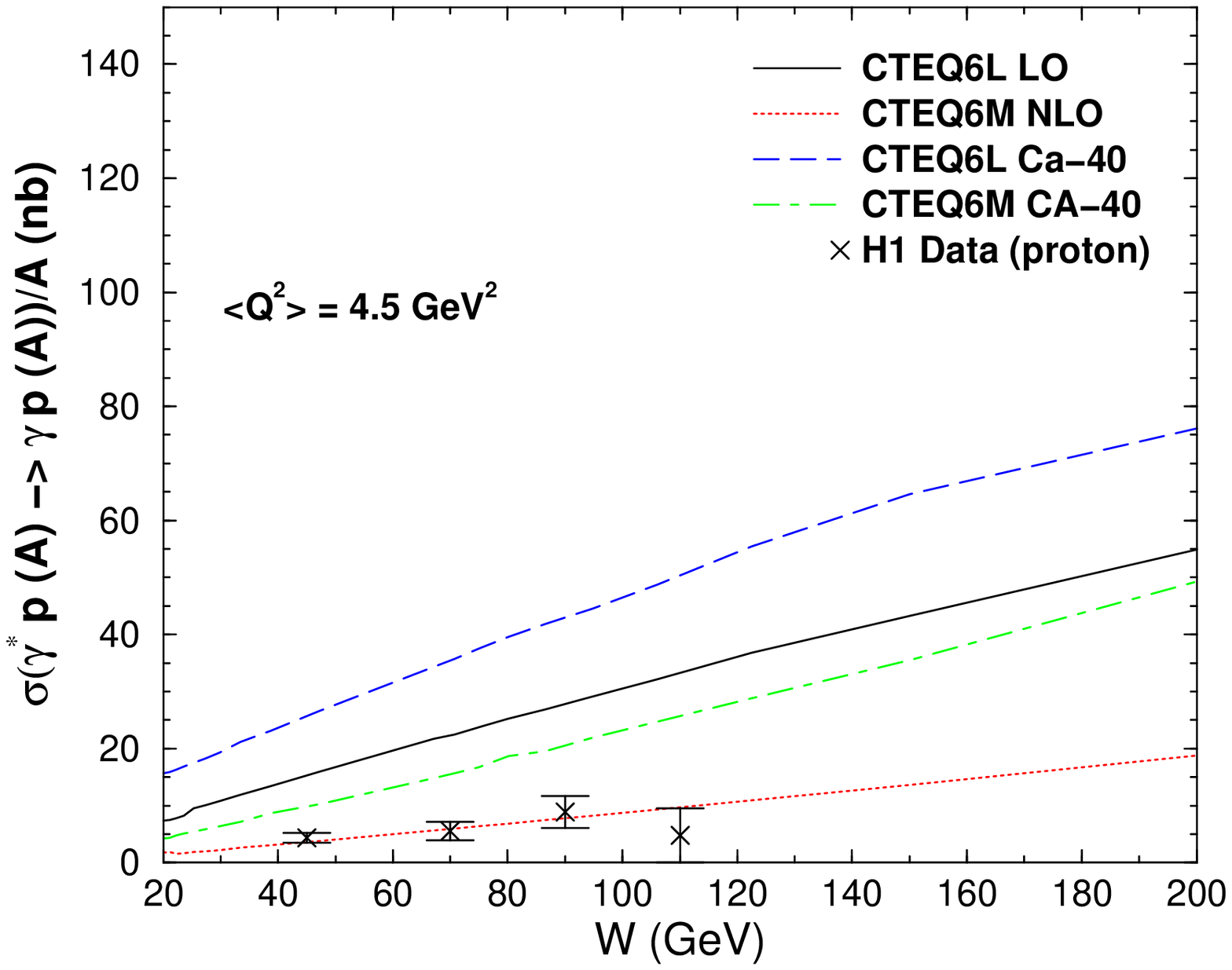,width=8.5cm,height=10.5cm}} 
\caption{One-photon cross section $\sigma(\gamma^*p)$ per nucleon for fixed $Q^2$ vs. $W$. H1 data on the nucleon is plotted in comparison within the same kinematics.}
\label{sigwca-40}
\end{figure}

\begin{figure}  
\centering
\mbox{\epsfig{file=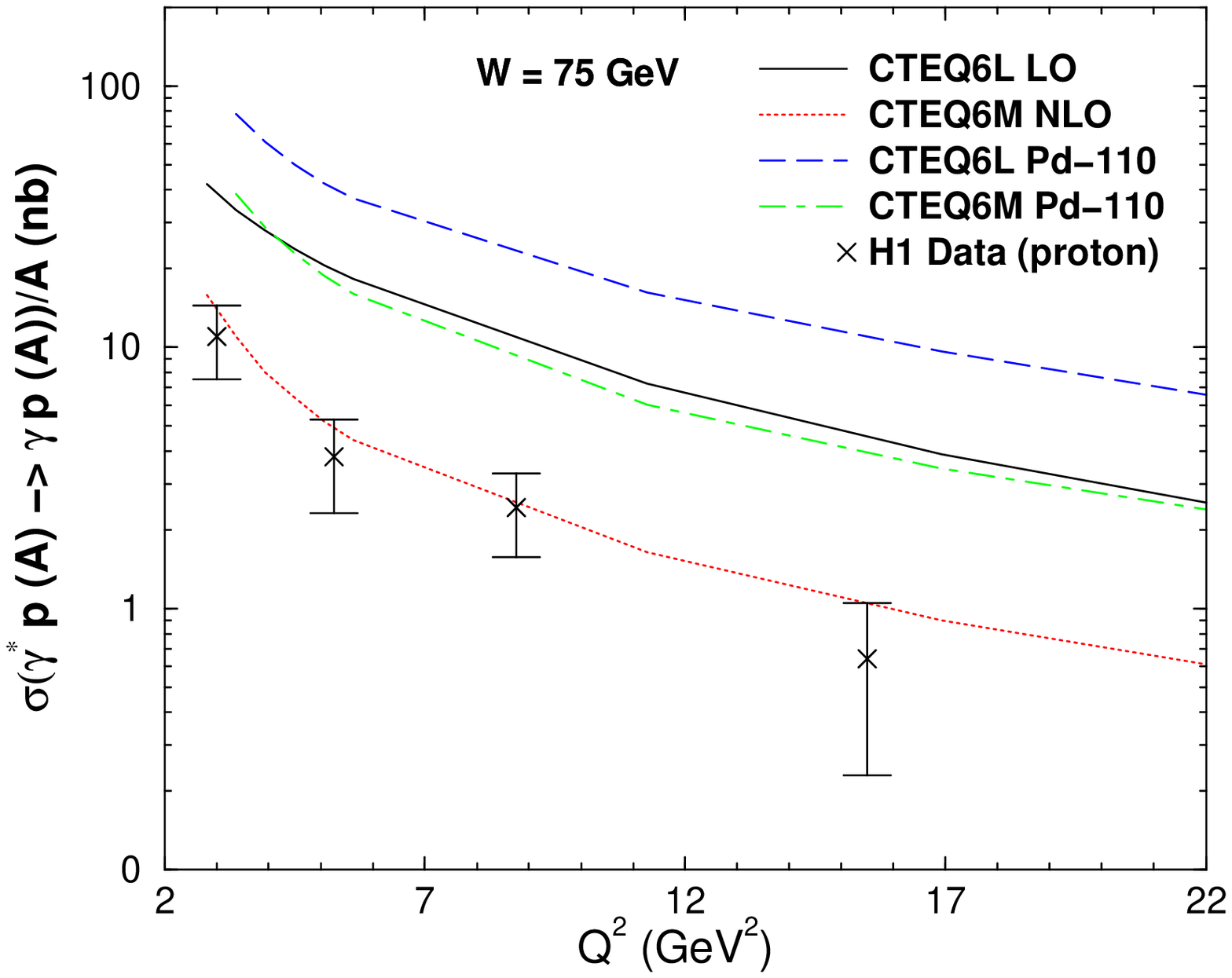,width=8.5cm,height=10.5cm}} 
\caption{One-photon cross section $\sigma(\gamma^*p)$ per nucleon for fixed $W$ vs. $Q^2$. H1 data on the nucleon is plotted in comparison within the same kinematics.}
\label{sigqpd-110}
\vskip+0.2in
\mbox{\epsfig{file=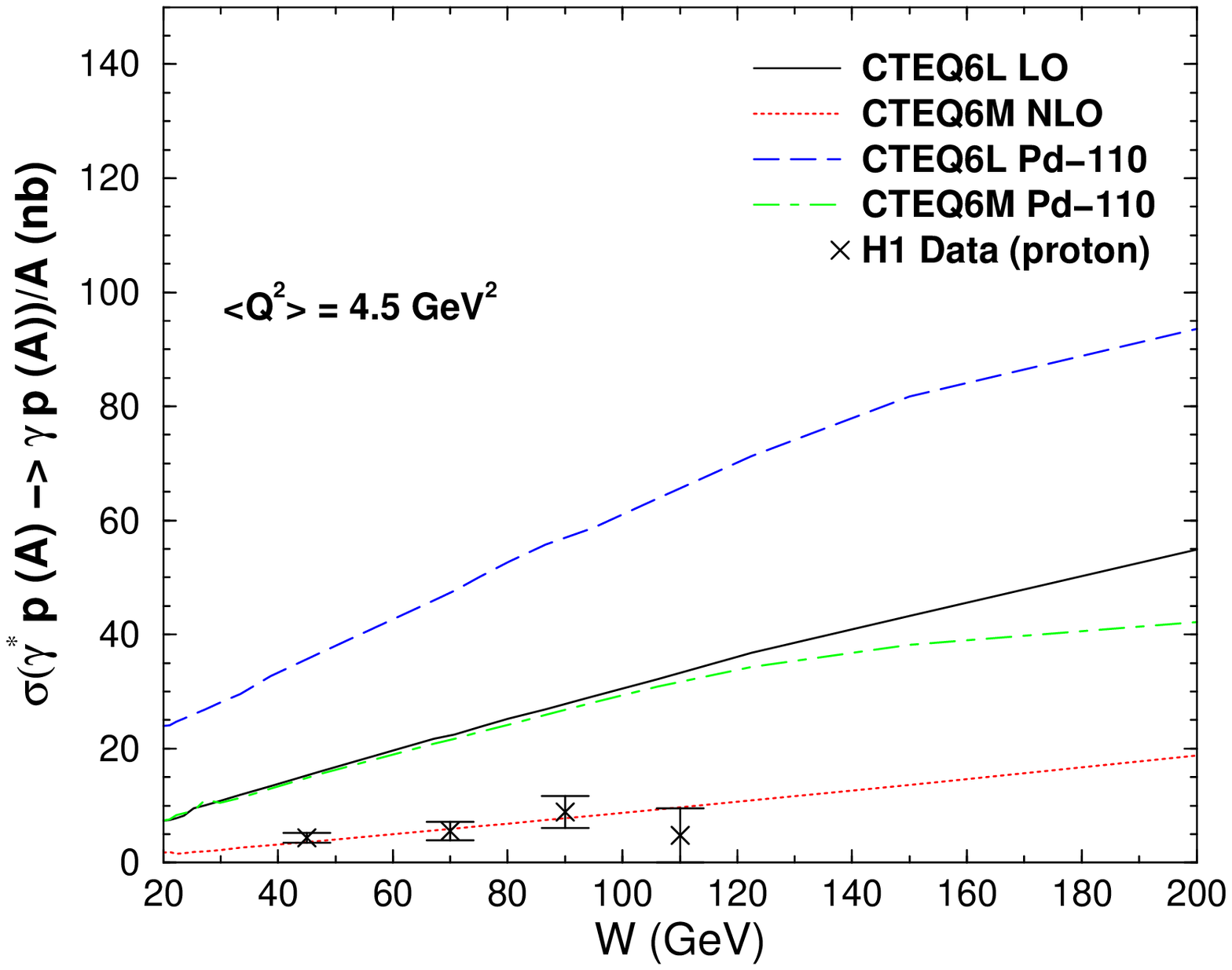,width=8.5cm,height=10.5cm}} 
\caption{One-photon cross section $\sigma(\gamma^*p)$ per nucleon for fixed $Q^2$ vs. $W$. H1 data on the nucleon is plotted in comparison within the same kinematics.}
\label{sigwpd-110}
\end{figure}

\begin{figure}  
\centering
\mbox{\epsfig{file=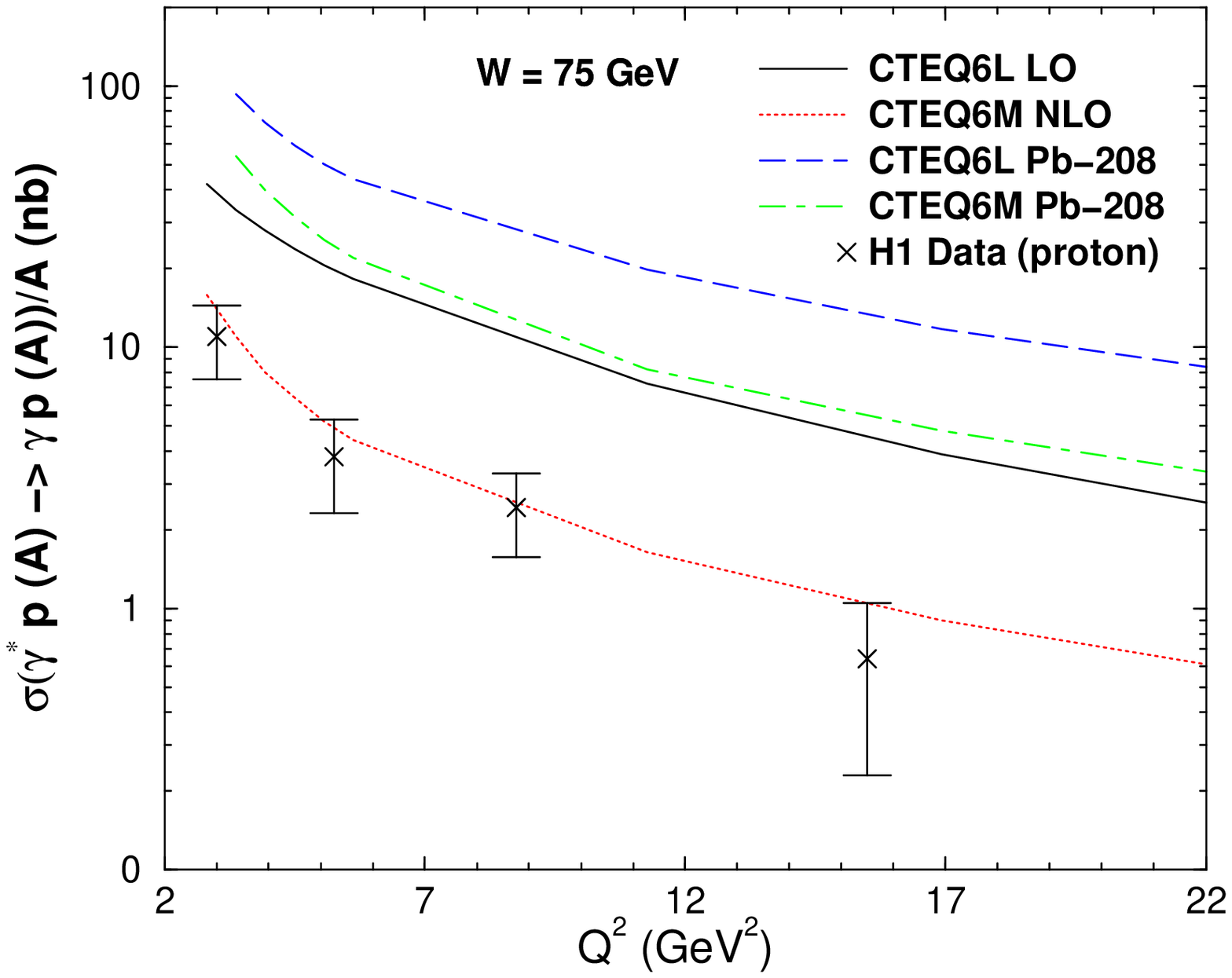,width=8.5cm,height=10.5cm}} 
\caption{One-photon cross section $\sigma(\gamma^*p)$ per nucleon for fixed $W$ vs. $Q^2$. H1 data on the nucleon is plotted in comparison within the same kinematics.}
\label{sigqpb-206}
\vskip+0.2in
\mbox{\epsfig{file=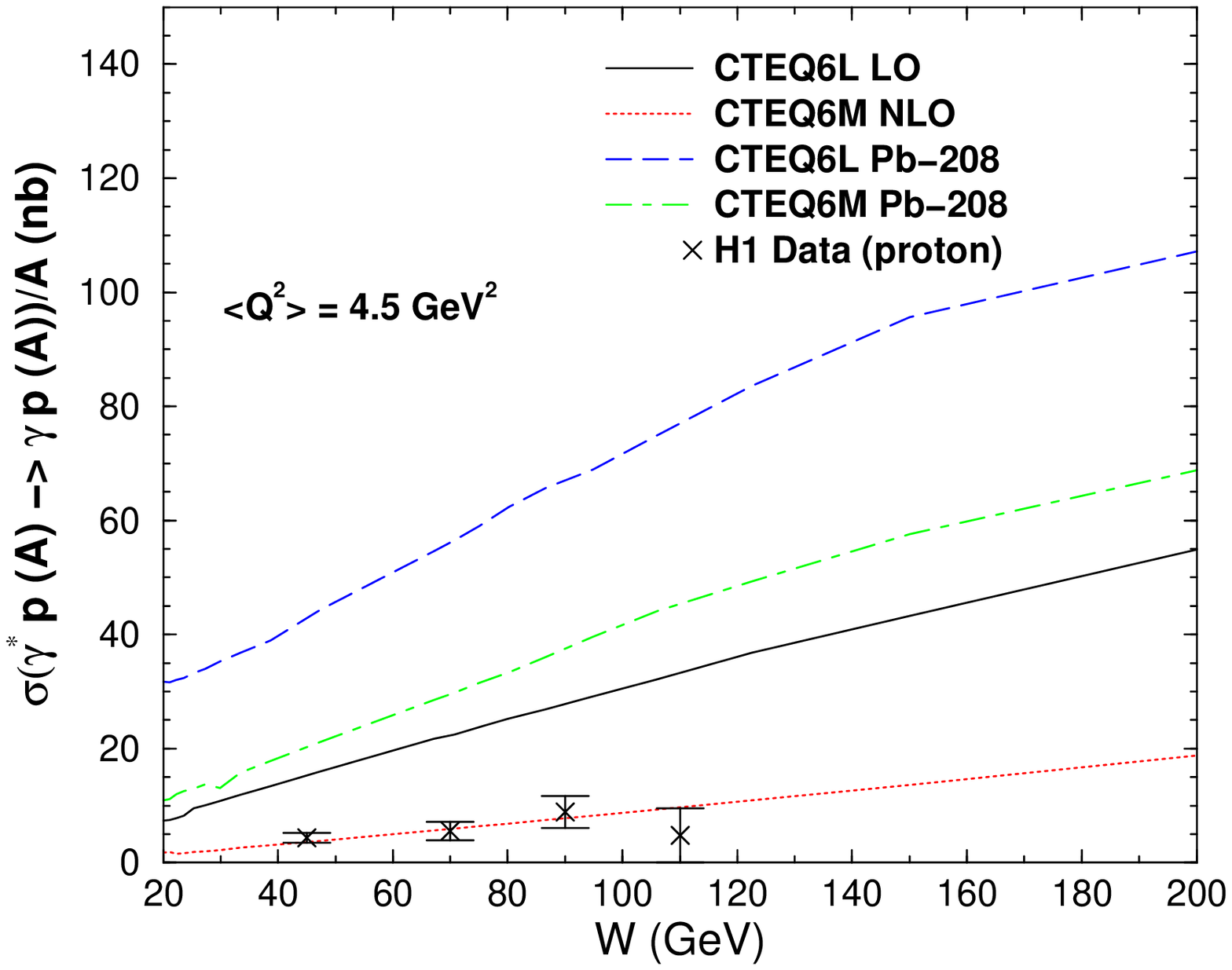,width=8.5cm,height=10.5cm}} 
\caption{One-photon cross section $\sigma(\gamma^*p)$ per nucleon for fixed $Q^2$ vs. $W$. H1 data on the nucleon is plotted in comparison within the same kinematics.}
\label{sigwpb-206}
\end{figure}

The one-photon cross section of DVCS per nucleon strongly increases
with $A$ even in the region where the nuclear shadowing effects are
important, see Figs.  \ref{sigqo-16} -- \ref{sigwpb-206}. As can be
seen from the figures, the relative LO to NLO correction is less
compared to the nucleon case and in line with the observation made for
the SSA and AAA. The main problem for the measurement of the inclusive
DVCS cross section is the subtraction of the BH QED background which
also increases with A.  The ratio, $R$, of the DVCS to BH cross
section is above $1$ for $\Bx\geq0.007$ and $Q^2=3~\mbox{GeV}^2$ and
rapidly increases as $\Bx$ increases ($R\simeq10^4$ for $\Bx\geq0.1$). This is
due to the fact that at fixed $Q^2$, $y$ deceases as $\Bx$ increases
and, hence, BH which is $\propto y^2$, dies out rapidly. As $Q^2$ increases
the value of $\Bx$ where $R\geq1$ increases as well since $y$ now
increases. For example, at $Q^2=25~\mbox{GeV}^2$, $R\geq1$ for
$\Bx\geq0.06$. These statements are essentially independent of $A$ and
imply a very broad kinematic window where DVCS dominates over BH,
making a measurement of $\sigma(\gamma^*A)$ on the $10\%$ level after subtraction
of the BH contribution very feasible.  Hence, it would be possible to
perform two independent measurements of the imaginary part of the
amplitude in a rather wide range of $\Bx$, $Q^2$, since $\sigma(\gamma^*A)$ is
dominated by the imaginary part of the nuclear DVCS amplitude.

\section{Conclusions}
We have demonstrated that experiments at the EIC will allow one to
measure the imaginary part of the nuclear DVCS amplitude using two
independent techniques, the SSA and $\sigma(\gamma^*A)$. We also predict a large
azimuthal angle asymmetry which could be studied using a high
resolution detector. Such experiments will enable us to obtain unique
information about quark nuclear GPDs, through scaling violations, as
well as about gluon nuclear GPDs. The study of the A-dependence of the
DVCS amplitude will allow one to determine the nuclear modification of
nuclear GPDs as a function of the nuclear thickness up to the
thicknesses $\rho_0\sim 2\cdot R_A\sim14~\mbox{fm}$ where $\rho_0$ is the average
nuclear density. Furthermore, it will allow us to verify the predicted
larger nuclear modifications of the nGPDs \cite{AFMS} i.e. the
modification of the three dimensional distribution and their
connection to leading twist nuclear shadowing in the respective
forward nuclear parton densities. These forward densities will also be
studied at the EIC. Verifying the expectations of large nuclear
effects for the real part of the DVCS amplitude at $x\sim0.1$ would
require running EIC at lower energies (which is also necessary for the
measurements of the longitudinal DIS cross section) such that BH will
be the dominating contribution in the cross section in the whole
kinematic range.

The information obtained from studies of nuclear DVCS will also enable
us to investigate the expected pattern of perturbative color opacity
in the coherent production of vector mesons on nuclei \cite{Brodsky}
in DIS at small $\Bx$, in an essentially model independent way.
 
\section*{Acknowledgment}
We thank Vadim Guzey for providing the numerical results for the
nuclear PDFs.  This work was supported by the DFG under the
Emmi-Noether grant FR-1524/1-3 and the DOE under grant number
DE-FG02-93ER40771.  MS's research was also supported by the
A.v.Humbold foundation.

\end{document}